\documentclass[prb,superscriptaddress,twocolumn,aps,showpacs]{revtex4}
\usepackage{amssymb}
\usepackage{graphicx}
\usepackage{longtable}
\usepackage{amsmath}
\usepackage{amsfonts}
\usepackage{color}
\usepackage{epsfig}
\usepackage{subfigure}

\begin{document}
\author{S. Carretta}
\affiliation{Dipartimento di Fisica e Scienze della Terra, Universit\`a di Parma, I-43124 Parma, Italy}
\author{A. Chiesa}
\affiliation{Dipartimento di Fisica e Scienze della Terra, Universit\`a di Parma, I-43124 Parma, Italy}
\author{F. Troiani}
\affiliation{S3 Istituto Nanoscienze, Consiglio Nazionale delle Ricerche, I-41100 Modena, Italy}
\author{D. Gerace}
\affiliation{Dipartimento di Fisica, Universit\`a di Pavia, via Bassi 6, I-27100 Pavia, Italy}
\author{G. Amoretti}
\affiliation{Dipartimento di Fisica e Scienze della Terra, Universit\`a di Parma, I-43124 Parma, Italy}
\author{P. Santini}
\affiliation{Dipartimento di Fisica e Scienze della Terra, Universit\`a di Parma, I-43124 Parma, Italy}
\date{\today }
\title{Quantum-Information Processing with Hybrid Spin-Photon Qubit Encoding}

\begin{abstract}
We introduce a scheme to perform quantum-information processing that is based on a hybrid spin-photon qubit encoding. The proposed qubits consist of spin-ensembles coherently coupled to microwave photons in coplanar waveguide resonators. The quantum gates are performed solely by shifting the resonance frequencies of the resonators on a ns timescale. An additional cavity containing a Cooper-pair box is exploited
as an auxiliary degree of freedom to implement two-qubit gates.
The generality of the scheme allows its potential implementation with a wide class of spin systems.
\end{abstract}
\maketitle

\vskip 3 cm

A classical computer is made of a variety of physical components specialized for different tasks.
In the same way, a quantum computer will probably be a hybrid device exploiting the best characteristics of
distinct physical systems. In this spirit, much work has recently been done to achieve strong
coupling of high-quality factor coplanar-waveguide resonators with superconducting qubits,
such as Cooper-pair boxes (CPBs) [\onlinecite{Wallraff,Majer,charge}] and transmons [\onlinecite{koch}] and/or spin ensembles (SEs) [\onlinecite{KuboPRL,Schuster}].
Superconducting qubits coupled to a microwave cavity field were proposed for quantum information
processing (QIP) [\onlinecite{CPB, phaseQ, 2qubitPhase}], using classical fields [\onlinecite{CPB}] or external voltages [\onlinecite{ghosh}] as a manipulation tool.
During the last years several theoretical works have considered the possibility of joining the fast processing
of superconducting qubits to the long coherence times of SEs [\onlinecite{Rabl, Imamoglu, Verdu, Wesenberg, review}], which
can be naturally exploited as quantum memories.
Cavity photons can be used as a bus to transfer the quantum state from CPBs to spin ensembles, and to couple distant CPB qubits,
leading to an effective interaction necessary to perform two-qubit gates [\onlinecite{wallquist}].
Recently, it was theoretically shown that a minimal architecture solely based on SEs can be exploited for full QIP  [\onlinecite{Benjamin}], by employing a measurement-based
scheme in which photons are still used as a quantum bus. \\
\begin{figure}
   \centering
   \includegraphics[width=7.25cm]{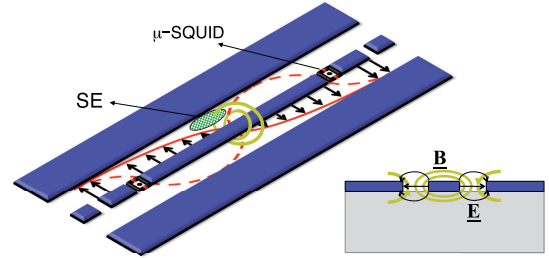}
   \caption{(Color online) Sketch of a tunable coplanar superconducting resonator in a microstrip line.
The effective resonator length can be tuned by inductively coupled micro-SQUIDs.
The fundamental (third) harmonic of the resonator is schematically shown with a solid
(dashed) line. The cavity-SE coupling is maximized at the magnetic field antinodes (rotational lines). Inset: cross-section of the strip lines on an insulating substrate.
}
   \label{CPW}
\end{figure}
Here we introduce a qualitatively different approach based on a hybrid spin-photon encoding of the qubits. Our scheme differs from previous ones because here the spin and photon degrees of freedom enter on an equal footing in the definition of the qubit.
This allows us to to perform all the manipulations simply by tuning the resonance frequencies of microstrip-line superconducting resonators (Fig.~1). 
Experimentally, the resonator frequencies have already been shown to be variable on a ns timescale [\onlinecite{KuboPRA}],
and up to tenths of the fundamental-mode frequency $\omega_c^0$ [\onlinecite{tunability}].
Besides, they can be assembled in a series of lumped elements, realizing large arrays of different
geometries [\onlinecite{houck_review}].
Such capability, along with the possibility of individually addressing the resonators, represents a prerequisite for scalability.
As to the spins, the few requisites of the scheme can be fulfilled by a large variety of systems, ranging from diluted transition-metal or rare-earth ions
to molecular nanomagnets [\onlinecite{Gatteschibook}].
In our hybrid encoding, each physical qubit is represented by a resonator mode and a SE. We describe the quantum gates in the elementary unit of a scalable setup, namely two resonators containing different SEs, where each qubit is encoded in the state of a distinct SE-resonator pair. These are connected by an interposed cavity containing a CPB, which is exploited as an auxiliary degree of freedom to implement two-qubits gates. \\
{\it Definition of the qubit --} We consider a resonant cavity containing a single photon in a mode of frequency $\omega_c$, and an ensemble of
$N$ identical and non-interacting spins $1/2$, initially prepared in the ground state $\left|\psi_0\right\rangle = \left|0...0\right\rangle$.
If the resonator mode is tuned to the gap of the two-level system, $\omega_1$, after some time  the SE will collectively absorb the
photon and evolve into the state $\left|\psi_1\right\rangle = \frac{1}{\sqrt{N}} \sum_{q=1}^{N} \left|0_1 ... 1_q ... 0_{N} \right\rangle$.
Transitions between $ \left|\psi_0\right\rangle $ and $ \left|\psi_1\right\rangle $ are described by $\hat{b}_1=\frac{1}{\sqrt{N}} \sum_{q=1,N} | 0 \rangle\langle 1 |_q $ and $\hat{b}_1^\dagger$ [\onlinecite{Wesenberg,KuboPRL}].
In the low-excitation regime, the collective excitations of the SE can be described as a harmonic oscillator and $ [ \hat{b}_1, \hat{b}_1^{\dagger} ] = 1 $.\\
Within the single-excitation subspace of the system formed by a resonator mode and a SE, we introduce the hybrid encoding of the qubit $\mu$:
\begin{equation}\label{encoding}
| 0 \rangle_\mu \equiv | \psi_1^\mu , n_\mu = 0 \rangle , \
| 1 \rangle_\mu \equiv | \psi_0^\mu , n_\mu = 1 \rangle ,
\end{equation}
where $ n_\mu $ 
is the photon occupation number of the cavity mode coupled with the SE.
Thus, the logical state of the qubit depends on whether the excitation is stored within the SE
or the quantized field of the resonator.\\
{\it Description of the scalable setup--}
To achieve universal QIP it is sufficient to perform a two-qubit quantum gate such as the controlled$-Z$ (CZ),
and arbitrary single-qubit rotations around two non-parallel axes [\onlinecite{Nielsen}].
Hereafter, we describe our scheme for the quantum-gate implementation in the basic unit of a scalable setup, i.e., a system of two qubits ($\mu =A, A'$), encoded in the hybrid states of two distinguishable SEs coupled to the modes of two different stripline resonators, as in Fig.~2(a).
In its general form, the scheme can be implemented within any bipartite lattice of cavities [\onlinecite{array}].
Both SEs consist of effective $s=1/2$ spins, but with different energy gaps: $\omega^A$ is coupled to the harmonic of frequency $\omega_c^A$ in cavity $A$, while $\omega^{A'}$ to the harmonic of frequency $\omega_c^{A'}$ in cavity $A'$, as in the level scheme of Fig.~2(b). A third resonator $B$, which is not used to encode qubits, is located in between the cavities $A$ and $A'$. It contains a CPB which we exploit to implement CZ. As we show in Appendix A, for reasonable values of the Josephson and charge energies [\onlinecite{Wallraff}], the CPB is characterized by the anharmonic spectrum reported in the central part of Fig.~2(b) (hereafter we consider only the three lowest levels). In cavity B we consider two different harmonics, $\omega_c^{B}$ and $\omega_c^{B'}$, respectively close to the gaps $\omega^B$ and $\omega^{B'}$ of the CPB.
\begin{figure}
   \centering
   \includegraphics[width=8.3cm]{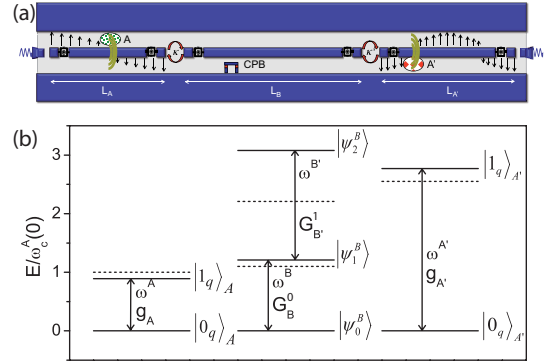}
   \caption{(Color online)
   (a) Representation of an elementary unit of the scalable setup.
   (b) Level diagram (solid line) of the two spin systems used to define the qubits and of the interposed CPB, which is used to implement CZ gates.
   The individual spin-photon
   coupling strengths are indicated as $g_{\gamma}$, corresponding to transition frequency $\omega^\gamma$ between single-spin states $\left| 0_q \right\rangle_\gamma$ and $\left| 1_q
   \right\rangle_\gamma$ ($\gamma =A,A'$). The CPB-cavity couplings are indicated with capital letters. Dashed lines represent the frequencies of the cavity modes.}
   \label{1cavity}
\end{figure}
In the idle configuration, the cavity modes $A-B$ and $B'-A'$ are detuned and qubits encoded in different cavities evolve independently from one another.
Hence, single qubit rotations can be implemented
by  varying the resonance frequency of the relevant resonator, while the CPB is left in its ground state and no photons are present in cavity $B$. The interaction between
neighboring qubits is switched on when the two cavities $A$ and $A'$ are brought into resonance with $B$ and photons can jump from cavities $A$ and $A'$ to $B$.
As explained below, CZ gates can be obtained by exploiting a two-step Rabi oscillation of the CPB between its ground state $|\psi_0^B\rangle$ and the excited state $|\psi_2^B\rangle$ (Fig.~2(b)). A three-level superconducting system has been exploited for performing CZ gates in a different scheme in [\onlinecite{Mariantoni}].\\
The total Hamiltonian of the SEs+CPB coupled to the quantized cavity fields reads:
\begin{equation}
\hat{H} = \hat{H}_{spin} + \hat{H}_{CPB} + \hat{H}_{ph}   + \hat{H}_{int} + \hat{H}_{hop} \, .
\label{hamil}
\end{equation}
The first term describes the SEs, $A$ and $A'$, as independent
harmonic oscillators [\onlinecite{KuboPRA}] ($\hbar=1$):
\begin{equation}
\hat{H}_{spin} = \omega^A \hat{b}^{\dagger}_{A } \hat{b}_{A }
                                                                + \omega^{A'}  \hat{b}^{\dagger}_{A' } \hat{b}_{A' } \,
\label{hamil2}
\end{equation}
The time-dependent photonic Hamiltonian, which in our scheme is entirely responsible for the qubits manipulation, is:
\begin{equation}
\hat{H}_{ph} = \sum_{\gamma=A,A',B,B'} \omega_c^\gamma (t) \hat{a}^{\dagger}_\gamma \hat{a}_\gamma \, ,
\label{hamil3}
\end{equation}
where $ \omega_c^\gamma (t) = \omega_c^\gamma (0) + \Delta_c^\gamma (t) $.
Within the rotating-wave approximation, the spin-photon and CPB-photon couplings read:
\begin{equation}
\hat{H}_{int} = \! \! \! \sum_{\gamma=A,A'} \! \!
\frac{\bar{G}_{\gamma}}{2}  \hat{a}^{\dagger}_\gamma \hat{b}_\gamma + \! \! \! \sum_{ \substack{
\gamma=B,B'\\
j=0,1}} \!
\frac{G_{\gamma}^j}{2}  \hat{a}^{\dagger}_\gamma |\psi_j^B \rangle \langle \psi_{j+1}^B | + \text{h.c.} \,  \label{hamil4}
\end{equation}
The coupling constants $\bar{G}_{\gamma}$ are enhanced by a factor $\sqrt{N}$ with respect to their
single-spin counterpart $g_{\gamma}$ (see, e.g., Ref. \onlinecite{Wesenberg}), while $G_\gamma^j$ are referred to a single superconducting unit.
Finally, the last term in  Eq. \ref{hamil} describes the photon hopping induced by the evanescent coupling
between the modes $A-B$ and $A'-B'$ of the two neighboring cavities [\onlinecite{evanescent,houck}]:
\begin{equation}
\hat{H}_{hop} = - \kappa \hat{a}^{\dagger}_{A} \hat{a}_{B}
                            - \kappa' \hat{a}^{\dagger}_{A'} \hat{a}_{B'} + \text{h.c.}
\label{hopping}
\end{equation}
Other hopping terms can be easily made negligible by engineering the two different
cavities. Hereafter, we will use the interaction picture, with $ \hat{H}_0 = \hat{H}_{spin}+ \hat{H}_{CPB} +\hat{H}_{ph}(t=0)$.\\
{\it Quantum gates--} In order to perform one- and two-qubit gates, we exploit the absorption (emission) of the photons entering the hybrid encoding (Eq. \ref{encoding}).
These processes can be straightforwardly controlled by tuning the frequencies of the cavity modes by a quantity $\Delta_c^\gamma$ for suitable time intervals. We will refer to such variations of the resonator frequencies as \textit{pulses}. In order to make the manipulation experimentally easier, we choose (see Fig.~2(b)) $\omega_c^B(0)$ to be intermediate between $\omega_c^A(0)$ and $\omega^B$, while $\omega_c^{A'}(0)$ is close to $\omega_c^{B'}(0)$ and $\omega_c^{B'}(0)$ is close to $\omega^{B'}$.
In the idle configuration $ \Delta_c^\gamma (t) = 0 $, the cavity-mode frequencies are significantly detuned from
the spin energy gaps ($ | \omega^\gamma_c (0) - \omega^{\gamma} | \gg \bar{G}_{\gamma} $) and $\hat{H}_{int}$ is ineffective. In addition, the cavities $A$ and $A'$ are far-detuned from $B$ ($ | \omega_c^A - \omega_c^{B} | \gg \kappa , | \omega_c^{A'} - \omega_c^{B'} | \gg \kappa'$)
and the effect of inter-cavity coupling is negligible. Hence, single-qubit gates can be performed independently on each cavity.\\
In particular, a rotation $\hat{R}_z(\varphi)$ of an arbitrary angle $\varphi$ about  the  $z$ axis of the Bloch sphere is performed by off-resonance pulses. These induce
a phase difference between the $|0\rangle_\mu$ and $|1\rangle_\mu$ states of the hybrid qubits (Eq. \ref{encoding}), given by $\int_{t_0-\tau/2}^{t_0+\tau/2} \Delta_c^\gamma(t) dt$ ($\tau$ is the pulse duration). For simplicity we assume step-like pulses,
$\Delta_c^\gamma(t) \equiv \, \delta_c^\gamma \, \theta(\tau/2-|t-t_0|)$, so that a generic $\hat{R}_z(\varphi)$ rotation is obtained by setting
$\delta_c^\gamma \tau = \varphi$.
Conversely,
$\hat{R}_y(\varphi)$ rotations are obtained by tuning the frequency of the cavity mode to match the corresponding energy gap of the SE
($ \delta^\gamma_c = \omega^\gamma - \omega^\gamma_c (0) $, with $\gamma = A,A'$) for the time $\tau = \varphi / \bar{G}_{\gamma} $.  The additional phase $\phi = \delta_c^\gamma \tau$ of the $\left|1\right\rangle_\mu$  qubit state is
straightforwardly corrected by an $\hat{R}_z(-\phi)$ rotation.
A simulation of this gate is reported in Fig.~3(a) in terms of the overlaps $ c_{i j} (t) = \langle i_A j_{A'} | \psi (t) \rangle $
between the system state $ | \psi (t) \rangle $ and the logical two-qubits states $|i_A j_{A'} \rangle =|i_A \rangle \otimes |j_{A'} \rangle$ ($i,j=0,1$).\\
The implementation of two-qubit gates requires the coupling between the degrees of freedom that are used to encode the two qubits.
The CZ gate is performed with a two-step Rabi oscillation of the CPB between $| \psi_0^B \rangle$ and $| \psi_2^B \rangle$ accompanied by the absorption and emission of the two photons entering the definition of the two qubits. A multi-step pulse sequence involving the auxiliary states
\begin{eqnarray}
\!\!\! \!\! | \eta \rangle & \!\!\!= \!\!\!& | \psi_0^A , n_A = 0 \rangle  | \psi_0^B , n_{B} = 1, n_{B'} = 1 \rangle  |\psi_0^{A'} n_{A'}=0\rangle
 \nonumber\\
\!\!\! \!\!  | \xi  \rangle & \!\!\!= \!\!\!&  | \psi_0^A , n_A = 0 \rangle  | \psi_1^B , n_{B} = 0, n_{B'} = 1 \rangle  |\psi_0^{A'} n_{A'}=0\rangle
\nonumber\\
\!\!\! \!\! | \zeta  \rangle & \!\!\!=\!\!\! &  | \psi_0^A , n_A = 0 \rangle  | \psi_2^B , n_{B} = 0, n_{B'} = 0 \rangle  |\psi_0^{A'} n_{A'}=0\rangle
\label{aux}
\end{eqnarray}
is adopted.
We illustrate it considering the two qubits initially in $|1_A 1_{A'}\rangle$. The first step corresponds to the transfer of the photon of mode $A$ in cavity $A$ into mode $B$ in cavity $B$, by means of a $\pi$-pulse that brings the two modes into resonance, $ \delta_{c}^{A} = \omega^B_c (0) - \omega^{A}_{c} (0) $. Simultaneously, the photon of mode $A'$ is transferred into mode $B'$ by varying $\delta_c^{A'}$. This induces
the transition $ |1_A 1_{A'}\rangle \longrightarrow | \eta  \rangle $. As a second step, the photon of frequency $\omega^B_c$ is absorbed by the CPB, after a $\pi$-pulse corresponding to $ \delta_{c}^{B} = \omega^B-\omega_c^B(0)$, bringing the system into state $| \xi\rangle$.
Then, a $2\pi$-pulse corresponding to $ \delta_{c}^{B'} = \omega^{B'}_c(0) - \omega^{B'}$
brings the mode $\omega^{B'}_c$ into resonance with the $ | \psi_1^B \rangle \longleftrightarrow | \psi_2^B \rangle $ transition of the CPB, thus inducing a complete Rabi flopping between the states $ | \xi \rangle $ and $ | \zeta \rangle $.
As a result, a phase $\pi$ is added to $ | \xi \rangle $.
Finally, the repetition of the first two steps brings the state back to $ |1_A 1_{A'}\rangle $ with an overall phase $\pi$ [\onlinecite{nota1}]. The time evolution of the two-qubit state $ | \psi (t) \rangle $ induced by this pulse sequence is reported in Fig.~3(b).
Conversely, the other basis states do not acquire any phase and therefore this sequence implements the CZ gate. Indeed, the basis states $|0_A 1_{A'}\rangle$ and $|1_A 0_{A'}\rangle$ contain only one photon and thus the full two-step Rabi oscillation cannot occur (see Appendix B), while the state $|0_A 0_{A'}\rangle$ is completely unaffected.\\
\begin{figure}
   \centering
   \includegraphics[width=8.3cm]{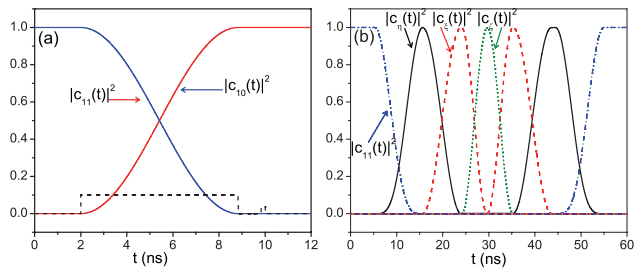}
   \caption{(Color online) (a) Calculated time-dependence of the components of $\left|\psi(t)\right\rangle$ in a
   single-qubit rotation $R_y(\pi)$. Dashed black lines represent $\Delta_c^\gamma (t)/\omega_c^\gamma(0)$. Negligible components are not shown.
   (b) Time-dependence of the main components of $\left|\psi(t)\right\rangle$ in a CZ gate. Auxiliary states $\left|\eta\right\rangle_{AA'}$, $\left|\xi\right\rangle_{AA'}$ and $\left|\zeta\right\rangle_{AA'}$ are described in Eq. \ref{aux}.}
   \label{2cavity}
\end{figure}
These simulations have been performed by assuming fundamental frequencies $\omega_c^{0A} / 2\pi=22$, $\omega_c^{0A'} / 2\pi=21$ and $\omega_c^{0B} /2\pi=12.5$
GHz [\onlinecite{Schuster}], and using the first and second harmonic (second and third harmonics) for the $A$ and $A'$ ($B$) cavities. With this choice, photon hopping between modes other than those included in  Eq. \ref{hopping} is negligible. As a figure of merit for CZ, we compute the \emph{fidelity-loss} $\lambda$ corresponding to the basis states
$ | i_A j_{A'} \rangle $:
\begin{equation}
\lambda \equiv \max_{i,j} \left\{1-\mathcal{F}^2_{ij}\right\} = \max_{i,j} \left\{ 1 -\left\langle i_A j_{A'} \left| \hat{\rho} \right| i_A j_{A'} \right\rangle \right\},
\label{fidelity}
\end{equation}
where $ \hat{\rho}=| \psi(t_f) \rangle \langle \psi(t_f) |$ is the final system state, when the qubits are initialized in $| i_A j_{A'} \rangle$.
We found $\lambda = 8 \times 10^{-4}$ for CZ and less for the other gates demonstrating the accuracy of the operations.
Analogous values of $\lambda$ are obtained for initial states corresponding to linear superpositions of the basis states.
Smaller values of $\lambda$ can be obtained by using larger frequencies or larger detunings.\\
The frequency is varied up to $\pm 0.1~\omega_c^{0}$ for the fundamental harmonic, and proportionally for the others.
Similar detunings have been experimentally shown in Ref. \onlinecite{tunability}.
We have assumed realistic values of the CPB-cavity  $G_{\gamma}^j / 2\pi = 60-90\ $MHz and SE-cavity coupling rates $\bar{G}_{\gamma} / 2\pi = 60\ $MHz, corresponding to $N \sim 10^{12}$ spins [\onlinecite{KuboPRL}], and tunneling rate $\kappa/ 2\pi=25\ $MHz, which has already been shown experimentally  [\onlinecite{houck}].
Larger values of $G_{\gamma}$ and $\kappa$ would reduce the gating times, but would also increase the fidelity-loss $\lambda$,
unless compensated by larger detunings.\\
Differently from alternative approaches (see e.g., [\onlinecite{feng}]), the main source of decoherence within our hybrid qubit
encoding is determined by photon leakage due to cavity
modes losses, since spin coherence times can be very
long (see below). We have checked that cavity loss rates
in the 10 kHz range [\onlinecite{1photon,cleland}], i.e. photon lifetime of
tens of microseconds, result only in few photons lost
over thousand within the gates timescale. We report
full numerical simulations including these effects in Appendix C, confirming the robustness of our scheme. We notice
that pure dephasing of cavity modes is negligible in such
resonators [\onlinecite{decoerenza}]. As far as the CPBs are concerned, in the present scheme they are always in the ground state, except during the implementation of CZ gates. Using a CPB as a non-linear element instead of a transmon or a phase qubit increases the anharmonicity of the system, while reducing the coherence times [\onlinecite{koch}]. However, it is sufficient that the coherence time of the CPB is much longer than the CZ gating time. This can be easily achieved with present technology [\onlinecite{deph}].
At last, the tuning of the resonator frequency could introduce extra dissipation to the system. These effects have been analyzed experimentally (see, e.g., Ref. [\onlinecite{tunability}]) and found to be small. They are mainly attributed to thermal noise and they could be further reduced with high frequency resonators, as the ones assumed in this work, and working at very low temperatures where Q-factors are known to be larger and more stable.\\
{\it Physical implementation --}
Initialization and read-out can be achieved by exploiting the CPBs [\onlinecite{1photon,reviewCPB}]. Several classes of systems can be exploited as SEs for the implementation of our scheme.
A simple possibility would be to exploit the different gyromagnetic factors of 3d and 4f ions, in order to obtain suitable level schemes in an applied magnetic field.
The embedding of these ions in a non-magnetic crystalline matrix allows to reduce harmful dipolar interactions, and to enhance the decoherence time up to the ms range [\onlinecite{ions}].
 Another promising class of spin systems is represented by molecular nanomagnets (MNMs) [\onlinecite{Gatteschibook}].
 These spin clusters form crystals in which the molecules behave as identical and
non-interacting magnetic units, and can be diluted by using non-magnetic
 analogues.
Besides, MNMs present degrees of freedom, such as spin chirality, that can be manipulated by the electric-field component of the cavity field [\onlinecite{Trif}]
and are substantially protected from the magnetic environment [\onlinecite{Troiani}].\\
In summary, we have developed a scheme for quantum-information processing with spin ensembles in
superconducting stripline resonators, exploiting a hybrid spin-photon encoding of the qubits.
Our scheme is qualitatively different from previous ones because the spin and photon degrees of freedom enter the definition of the qubit on an equal footing.
In this way, the evolution can be induced simply by tuning the cavity frequency to the spin-energy gaps.
Arbitrary single and two-qubit gates are implemented, over much shorter times than typical decoherence times
of cavity-photons and spin-ensembles.
Promising candidates for the spin degrees of freedom are diluted magnetic ions or molecular
nanomagnets. The application of this scheme to an $ABAB...$ array of cavities enables general quantum algorithms
as well as quantum simulators [\onlinecite{houck_review,simulator}] to be implemented.\\

Very useful discussions with A. Auff\`eves are gratefully acknowledged.
This work was financially supported by the Italian FIRB project
``New challenges in molecular nanomagnetism: from spin dynamics to quantum-information processing''
and by Fondazione Cariparma.

\appendix

\section{Cooper-pair boxes}
The scheme for quantum information processing described in the main text exploits a Cooper-pair box (CPB) [\onlinecite{reviewCPB,koch}], whose Hamiltonian can be written in the basis of the charge states $|n\rangle$ (with $\hat{n} |n\rangle = n |n\rangle$) [\onlinecite{charge}] as
\begin{eqnarray} \nonumber
\hat{H}_{CPB}&=&\sum_{n=-\infty}^\infty 4E_C (n-n_g)^2 |n\rangle \langle n| \\
&-& \sum_{n=-\infty}^\infty \frac{E_J}{2} \left( |n\rangle \langle n+1| + |n+1\rangle \langle n|\right),
\label{CPB2}
\end{eqnarray}
where $E_J$ is the Josephson energy, $E_C$ is the charging energy and $n_g$ is the dimensionless gate charge.
The numerical diagonalization of (\ref{CPB2}) with $n_g = 0.5$, $E_C = 4.9 ~GHz$ and $E_J = 6.2~E_C $ [\onlinecite{Wallraff}]
yields a highly anharmonic spectrum. Hence, we can safely truncate the Hilbert space of the CPB to the lowest three levels shown in the central part of Fig.~2(b) of the main text. In this regime coherence times of the CPB are considerably longer than the CZ gating time [\onlinecite{deph}] and the spectrum is sufficiently anharmonic to obtain high fidelities in our simulations.

\section{Proof-of-principle experiment}
We describe here a simpler setup, which could be exploited for the first
proof-of-principle experiments. This is formed by a CPB and two
distinguishable spin $s=1/2$ ensembles $A$ and $A'$ within the same cavity.
We exploit the lowest three levels of the CPB: $|\psi_0^B\rangle$, $|\psi_1^B\rangle$ and $|\psi_2^{B}\rangle$, with transition energies
$\omega^B$ and $\omega^{B'}$ ($\hbar=1$ is assumed). The excitation energies of the spin ensembles are $\omega^A$ and $\omega^{A'}$.
Two different harmonics of the resonator are taken into account, of frequency $\omega_c^A$ and $\omega_c^{A'}$. $\omega_c^A$ is intermediate between $\omega^A$ and $\omega^B$, while $\omega_c^{A'}$ is intermediate between $\omega^{A'}$ and $\omega^{B'}$. In this way the cavity mode $\omega_c^A$ ($\omega_c^{A'}$) can be coupled both to the spin gap $\omega^A$ ($\omega^{A'}$) and to the superconducting gap $\omega^B$ ($\omega^{B'}$), by tuning its frequency.
We recall here the definition of the logical 2-qubit states:
\begin{eqnarray} \nonumber
|0_A 0_{A'}\rangle \equiv | \psi_1^A, n_A=0 \rangle \otimes | \psi_1^{A'}, n_{A'}=0 \rangle \\ \nonumber
|0_A 1_{A'}\rangle \equiv | \psi_1^A, n_A=0 \rangle \otimes | \psi_0^{A'}, n_{A'}=1 \rangle \\ \nonumber
|1_A 0_{A'}\rangle \equiv | \psi_0^A, n_A=1 \rangle \otimes | \psi_1^{A'}, n_{A'}=0 \rangle \\
|1_A 1_{A'}\rangle \equiv | \psi_0^A, n_A=1 \rangle \otimes | \psi_0^{A'}, n_{A'}=1 \rangle
\label{logical}
\end{eqnarray}
We note that the CPB does not enter in the definition of the qubits: in the idle configuration it remains in its ground state $|\psi_0^B\rangle$.

Single-qubit rotations are implemented in a way similar
to that described in the text: rotations about the $z$ axis of the Bloch sphere, $\hat{R}_z(\varphi)$, are realized by means of off-resonant pulses, while $\hat{R}_y(\varphi)$
are performed by tuning $\omega_c^\gamma$ to $\omega^\gamma$ for the proper amount of time ($\gamma=A, A'$).

By exploiting the hybrid encoding (\ref{logical}) the CZ gate can be implemented as in the scalable setup. Here, however,
a single resonator is considered and thus photon-hopping does not occur.
As stated in the text, CZ is performed with
a two-step Rabi oscillation of the CPB between $|\psi_0^B\rangle$ and $|\psi_2^{B}\rangle$ accompanied by the absorption and emission of the
two photons entering the definition of the two qubits. This is done through the intermediate states:
\begin{eqnarray} \nonumber
|\eta\rangle \equiv |\psi_0^A, n_A=0 \rangle \otimes |\psi_1^B \rangle \otimes | \psi_0^{A'}, n_{A'}=1 \rangle \\
|\xi\rangle \equiv |\psi_0^A, n_A=0 \rangle \otimes |\psi_2^B \rangle \otimes | \psi_0^{A'}, n_{A'}=0 \rangle.
\label{aux2}
\end{eqnarray}
\begin{figure}
   \centering
   \includegraphics[width=8.0cm]{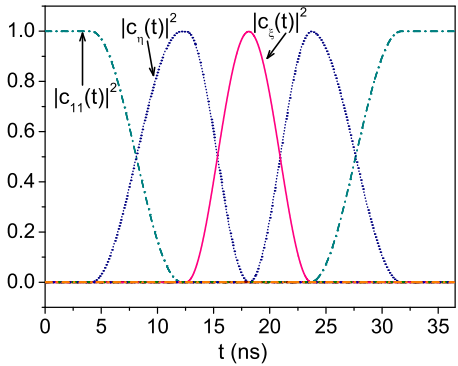}
   \caption{(Color online) Calculated time-dependence of the components of $\left|\psi(t)\right\rangle$ during the simulation of a CZ in the simpler setup described for the first proof-of-principle experiments.}
   \label{proof}
\end{figure}
We now examine the effect of the three-step pulse sequence used to implement CZ on the four logical states (\ref{logical}).
\begin{itemize}
\item Let's start by considering the two qubits initially prepared in state $|1_A 1_{A'}\rangle$. First step: $\omega_c^A$ is tuned to $\omega^B$ by means of a $\pi$-pulse, which transfers the excitation from the photon $A$ to the intermediate level $|\psi_1^B\rangle$ of the CPB, carrying the system to state $|\eta\rangle$. Second step: a $2 \pi$-pulse brings $\omega_c^{A'}$ into resonance with $\omega^{B'}$, thus inducing a full Rabi oscillation between states $|\eta\rangle$ and $|\xi\rangle$. Third step: the repetition of the first pulse brings back the system to $|1_A 1_{A'}\rangle$, with a phase $\pi$ acquired during the Rabi flop in the second step. The extra phase due to the absorption/emission processes occurring in the first and third steps can be set to zero by properly choosing the delay between the two corresponding $\pi$-pulses.
\item If the system is initially in $|1_A 0_{A'}\rangle$ there is only the photon of frequency $\omega_c^A$ and the $2 \pi$-pulse (second step) leaves the state $|\eta\rangle$ unaffected. The two $\pi$-pulses (first and third steps) brings the state from $|1_A 0_{A'}\rangle$ to $|\eta\rangle$ and back, but the phase acquired during this process is set to zero by the same delay fixed previously.
\item If the system is in $|0_{A} 1_{A'}\rangle$, the two $\pi$-pulses do not induce any transition. In fact, in this state there is only a photon in mode $A'$ which is never brought into resonance with the transition $\omega^B$. Thus, also the Rabi swap (second step) is ineffective, as the CPB is always in its ground state.
\item Finally, state $|0_A 0_{A'}\rangle$ is completely unaffected by the present sequence.

\end{itemize}
Hence, the sequence of pulses described above implements a CZ, i.e., in matrix representation on basis (\ref{logical}):
\begin{equation}   \nonumber
\left(
\begin{array}{cccc}
	~1 & ~0 & ~0 & ~0 \\
	~0 & ~1 & ~0 & ~0 \\
    ~0 & ~0 & ~1 & ~0 \\
    ~0 & ~0 & ~0 & -1 	
\end{array}
\right).
\end{equation}
As photon-hopping is not required, the gating times are reduced with respect to the scalable setup, to about $30~ns$ (see the simulation reported in Fig. \ref{proof}).

\section{Photon loss}
\begin{figure}
   \centering
   \subfigure{\includegraphics[width=4.1cm]{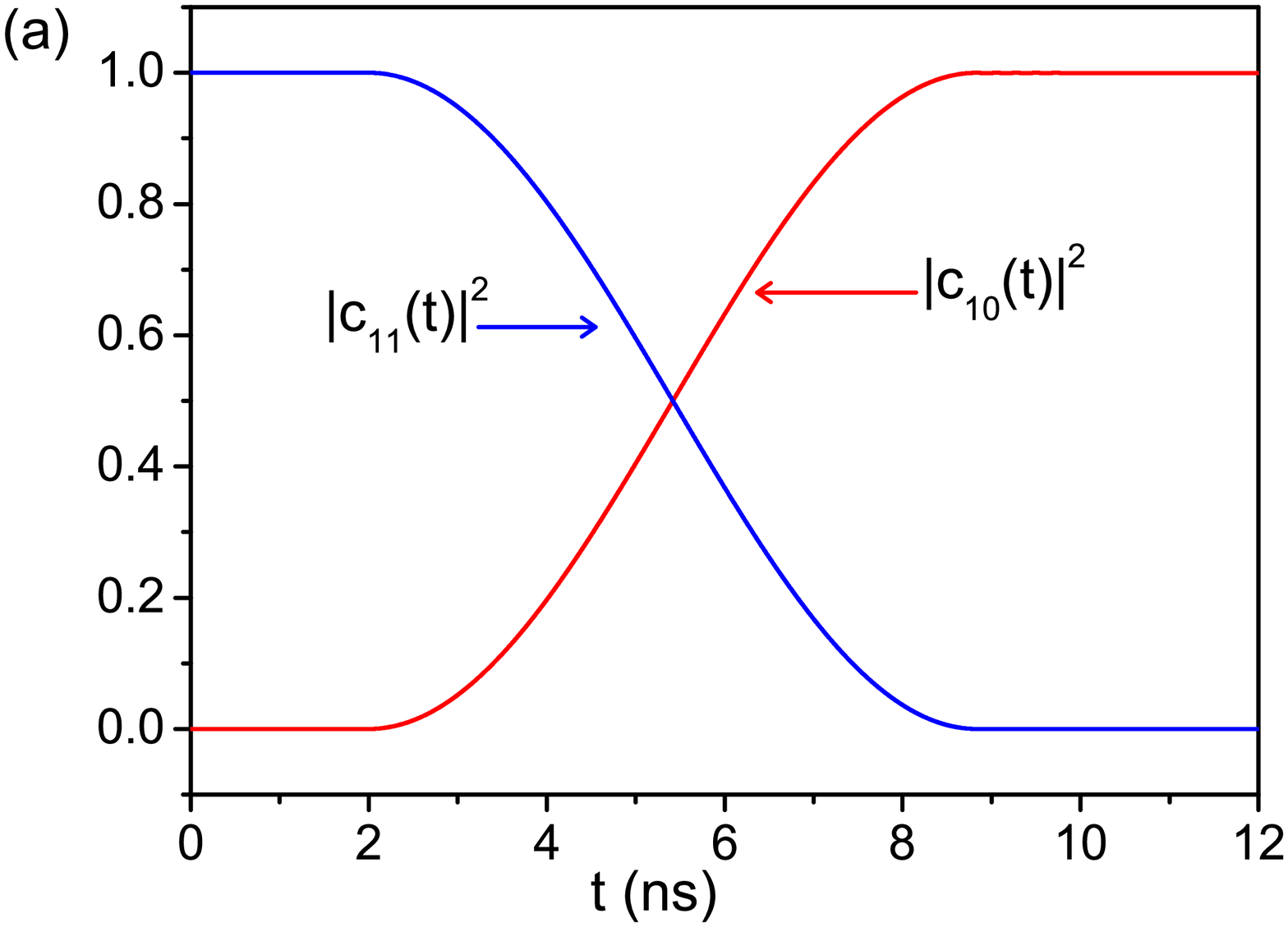}}
   \hspace{1mm}
   \subfigure{\includegraphics[width=4.1cm]{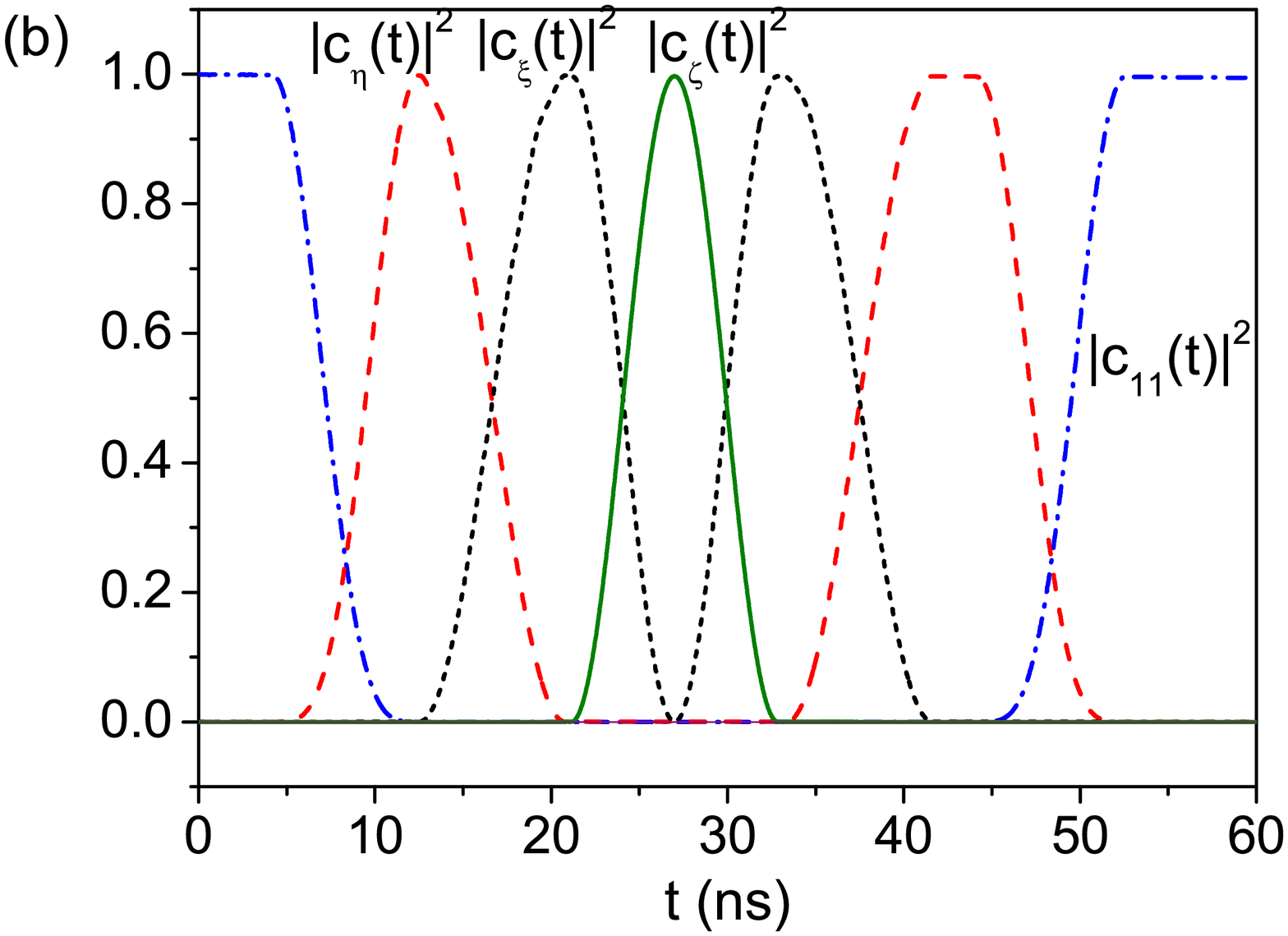}}
   \caption{(Color online) Calculated time-dependence of the main components of $\left|\psi(t)\right\rangle$ in a single-qubit rotation, (a), and in a CZ, (b), including a term of photon-leakage. Calculations are performed in Schr\"odinger picture.}
   \label{leakage}
\end{figure}

Following [\onlinecite{Auffeves}], we include the photon leakage by modeling the evolution of the system as due to an effective Hamiltonian
$H_{eff}$ involving the complex cavity frequencies $\tilde{\omega}_c^\gamma=\omega_c^\gamma-i \Gamma$. With this substitution
the photon part of the Hamiltonian changes into
$$
\sum_\gamma \omega_c^\gamma (t) \hat{a}_\gamma^{\dagger} \hat{a}_\gamma \rightarrow \sum_\gamma \left[ \omega_c^\gamma (t) - i \Gamma \right] \hat{a}_\gamma^{\dagger} \hat{a}_\gamma .
$$
Here $\Gamma$ is the cavity loss rate, assumed in the 10 kHz range [\onlinecite{1photon,cleland}], and $\omega_c^\gamma$ are the real frequencies introduced in the text.
Fig. \ref{leakage} shows the absolute squared value of the
components of the system wave-function as a function of time, during the simulation of a $\hat{R}_y (\pi)$ and of a CZ gate. No significant difference can be noted with respect to Fig. 3(b) of the text: the non-hermitian evolution due to $H_{eff}$ leads only to few photons lost over thousand during our gating times.
Finally, it has been experimentally shown that pure
dephasing of the cavity modes is negligible (see, e.g., [\onlinecite{decoerenza}] where the measured value of the dephasing time
$T_2$ approximately corresponds to twice the value of
the photon decay time $T_1$).

\section{Pulse shape}
In the simulations reported in the text the cavity modes are modulated by means of step-like pulses.
We have checked that the same results can be achieved by linearly varying the resonator frequency on a timescale
$\gg 1/\omega_c^\gamma (0)$: this regime is well within the validity of the rotating-wave approximation and no generation of unwanted photons due
to dynamical-Casimir effect is observed [\onlinecite{DCE2}].


\begin{thebibliography}{99}

\bibitem{Wallraff}
A.~Wallraff, D.I. Schuster, A.~Blais, L. Frunzio, R.~S. Huang, J.~Majer, S.~Kumar, S.M. Girvin, and R.J. Schoelkopf ,
Nature (London) {\bf 431}, 162 (2004); H. Paik, D. I. Schuster, Lev S. Bishop, G. Kirchmair, G. Catelani, A. P. Sears, B. R. Johnson, M. J. Reagor, L. Frunzio, L. I. Glazman, S. M. Girvin, M. H. Devoret, and R. J. Schoelkopf, Phys. Rev. Lett. {\bf 107}, 240501 (2011).

\bibitem{Majer}
J.~Majer, J.M.~Chow, J.M.~Gambetta, Jens Koch, B.R.~Johnson, J.A.~Schreier, L.~Frunzio, D.I.~Schuster,
A.A.~Houck, A.~Wallraff, A.~Blais, M.H.~Devoret, S.M.~Girvin, and R.J.~Schoelkopf,
Nature (London) {\bf 449}, 443 (2007).

\bibitem{charge}
Y. Makhlin, G. Sch\"on, A. Shnirman, Nature {\bf 398}, 305 (1999).

\bibitem{koch}
J. Koch, T.M. Yu, J. Gambetta, A.A. Houck, D.I. Schuster, J. Majer, A. Blais, M.H. Devoret, S.M. Girvin, and R.J. Schoelkopf,
Phys. Rev. A {\bf 76}, 042319 (2007).

\bibitem{KuboPRL}
Y.~Kubo, F.R.~Ong, P.~Bertet, D.~Vion, V.~Jacques, D.~Zheng, A.~Dr\'eau, J.-F.~Roch, A.~Auffeves,
F.~Jelezko, J.~Wrachtrup, M.F.~Barthe, P.~Bergonzo, and D.~Esteve,
Phys.~Rev.~Lett. {\bf 105}, 140502 (2010).

\bibitem{Schuster}
D.I.~Schuster, A.P.~Sears, E.~Ginossar, L.~DiCarlo, L.~Frunzio, J.J.L.~Morton, H.~Wu, G.A.D.~Briggs,
B.B.~Buckley, D.D.~Awschalom, and R.J.~Schoelkopf,
Phys.~Rev.~Lett. {\bf 105}, 140501 (2010).

\bibitem{CPB}
J.Q.~You, F.~Nori, Phys.~Rev.~B {\bf 68}, 064509 (2003);
A.~Blais, J.~Gambetta, A.~Wallraff, D.I.~Schuster, S.M.~Girvin, M.H.~Devoret, and R.J.~Schoelkopf,
Phys.~Rev.~A {\bf 75}, 032329 (2007).

\bibitem{phaseQ}
F. W. Strauch, P. R. Johnson, A.J. Dragt, C.J. Lobb, J. R. Anderson, F. C. Wellstood, Phy. Rev. Lett. {\bf 91}, 167005 (2003).

\bibitem{2qubitPhase}
L. DiCarlo, J.M. Chow, J.M. Gambetta, Lev S. Bishop, B.R. Johnson, D.I. Schuster, J. Majer, A. Blais,
L. Frunzio, S.M. Girvin, R.J. Schoelkopf, Nature {\bf 460}, 240 (2009).

\bibitem{ghosh}
J. Ghosh, A. Galiautdinov, Z. Zhou, A.N. Korotkov, J.M. Martinis, M.R. Geller, Phys. Rev. A {\bf 87}, 022309 (2013).

\bibitem{Rabl}
P.~Rabl, D.~DeMille, J.M.~Doyle, M.D.~Lukin, R.J.~Schoelkopf, and P.~Zoller,
Phys.~Rev.~Lett. {\bf 97}, 033003 (2006).

\bibitem{Imamoglu}
A.~Imamoglu, Phys.~Rev.~Lett. {\bf 102}, 083602 (2009).

\bibitem{Verdu}
J.~Verdu, H.~Zoubi, Ch.~Koller, J.~Majer, H.~Ritsch, and J.~Schmiedmayer,
Phys.~Rev.~Lett. {\bf 103}, 043603 (2009).

\bibitem{Wesenberg}
J.H.~Wesenberg, A.~Ardavan, G.A.D.~Briggs, J.J.L.~Morton, R.J.~Schoelkopf, D.I.~Schuster, and K.~Molmer,
Phys.~Rev.~Lett. {\bf 103}, 070502 (2009).

\bibitem{review}
Z.-L. Xiang, S.~Ashhab, J.Q.~You, and F.~Nori, Rev. Mod. Phys. (in press).

\bibitem{wallquist}
M.~Wallquist, V.S.~Shumeiko, and G.~Wendin,
Phys.~Rev.~B {\bf 74}, 224506 (2006).

\bibitem{Benjamin}
Y.~Ping, E.M.~Gauger, and S.C.~Benjamin,
New Journal of Physics, {\bf 14}, 013030 (2012).

\bibitem{KuboPRA}
Y.~Kubo, I.~Diniz, A.~Dewes, V.~Jacques, A.~Dr\'eau, J.-F.~Roch, A.~Auffeves, D.~Vion, D.~Esteve, and P.~Bertet,
Phys.~Rev.~A {\bf 85}, 012333 (2012).

\bibitem{tunability}
A.~Palacios-Laloy,  F.~Nguyen, F.~Mallet, P.~Bertet, D.~Vion, D.~Esteve, J.~Low~Temp.~Phys. {\bf 151}, 1034 (2008);
M.~Sandberg, F.~Persson, I.C.~Hoi, C.M.~Wilson, and P.~Delsing, Phys..~Scr., {\bf T137}, 014018 (2009).

\bibitem{houck_review}
A.A. Houck, H.E. Tureci, and J. Koch,
Nat.~Physics {\bf 8}, 292 (2012).

\bibitem{Gatteschibook}
D.~Gatteschi, R.~Sessoli, and J.~Villain,
\textit{Molecular nanomagnets}
(Oxford University Press, Oxford, UK, 2006).

\bibitem{Nielsen}
M.A. Nielsen and I.L. Chuang,
\textit{Quantum Computation and Quantum Information}
(Cambridge University Press, Cambridge, UK, 2000).

\bibitem{array}
J.-Q.~Liao, Z.R.~Gong, L.~Zhou, Y.-X.~Liu, C.P.~Sun, and F.~Nori,
Phys.~Rev.~A {\bf 81}, 042304 (2010).


\bibitem{deph}
J. Clarke, F.K. Wilhelm, Nature {\bf 453}, 1031 (2008).


\bibitem{Mariantoni} M. Mariantoni, H. Wang, T. Yamamoto, M. Neeley, R. C. Bialczak, Y. Chen, M. Lenander,
E. Lucero, A. D. O'Connell, D. Sank, M. Weides, J. Wenner, Y. Yin, J. Zhao, A. N. Korotkov,
A. N. Cleland, John M. Martinis, Science, {\bf 334}, 61 (2011).

\bibitem{evanescent}
M. Mariantoni, F. Deppe, A. Marx, R. Gross, F.K. Wilhelm, and E. Solano,
Phys.~Rev.~B {\bf 78}, 104508 (2008).

\bibitem{houck}
D.L. Underwood, W.E. Shanks, J. Koch, and A.A. Houck,
Phys.~Rev.~A {\bf 86}, 023837 (2012).

\bibitem{nota1} Additional trivial phases can be straightforwardly eliminated by short $\hat{R}_z$ operations as in the implementation of $\hat{R}_y$.
In this way also the phase acquired during photon-hopping processes can be zeroed. By properly setting the delay between the two $\pi$-pulses corresponding to $ \delta_{c}^{B} = \omega^B-\omega_c^B(0)$, the associated absorption/emission processes yields a zero additional phase.

\bibitem{feng} Z.B. Feng, Phys. Rev. A {\bf 85}, 014302 (2012).




\bibitem{1photon}
B.R.~Johnson,  M.D.~Reed,  A.A.~Houck, D.I.~Schuster, L.S.~Bishop, E.~Ginossar, J.M.~Gambetta,
L.~DiCarlo, L.~Frunzio, S.M.~Girvin, and R.J.~Schoelkopf,
Nat.~Physics {\bf 6}, 663 (2010).

\bibitem{cleland}
A. Megrant, C. Neill, R. Barends, B. Chiaro, Y. Chen, L. Feigl, J. Kelly, E. Lucero,
M. Mariantoni, P.J.J. O'Malley, D. Sank, A. Vainsencher, J. Wenner, T. C. White,
Y. Yin, J. Zhao, C. J. Palmstrom, J.M. Martinis, and A.N. Cleland,
Appl. Phys. Lett. {\bf 100}, 113510 (2012).

\bibitem{decoerenza} H. Wang, M. Hofheinz, M. Ansmann, R. C. Bialczak, E. Lucero, M. Neeley, A. D. O'Connell, D. Sank, J. Wenner,
A. N. Cleland, and John M. Martinis, Phys. Rev. Lett. {\bf 101}, 240401 (2008).


\bibitem{reviewCPB}
Y. Makhlin, G. Sch\"on, A. Shnirman, Review of Modern Physics {\bf 73}, 2 (2001).







\bibitem{ions} R. E. George, J. P. Edwards, and A. Ardavan, Phys. Rev. Lett. {\bf 110}, 027601 (2013).



\bibitem{Trif}
M.~Trif, F.~Troiani, D.~Stepanenko, and D.~Loss,
Phys. Rev. Lett. {\bf 101}, 217201 (2008).

\bibitem{Troiani}
F.~Troiani, D.~Stepanenko, and D.~Loss,
Phys. Rev. B {\bf 86}, 161409(R) (2012).

\bibitem{simulator}
P.~Santini, S.~Carretta, F.~Troiani, and G.~Amoretti,
Phys.~Rev.~Lett. {\bf 107}, 230502 (2011).

\bibitem{Auffeves}
I.~Diniz, S.~Portolan, R.~Ferreira, J.M.~G\'erard, P.~Bertet, and A.~Auff\'eves,
Phys.~Rev.~A {\bf 84}, 063810 (2011).

\bibitem{DCE2}
S. De Liberato,
D. Gerace, I. Carusotto, and C. Ciuti,
Phys.~Rev.~A {\bf 80}, 053810 (2009);
J. R.~Johansson, 
G.~Johansson, C.M.~Wilson, and F.~Nori,
Phys.~Rev.~Lett. {\bf 103}, 147003 (2009).

\end{thebibliography}
\end{document}